\documentclass[aps,prd,showpacs,preprintnumbers,nofootinbib,twocolumn]{revtex4-1}
\usepackage{bm}
\usepackage{latexsym}
\usepackage{dcolumn}
\usepackage{lipsum}
\usepackage{graphicx,epsfig,float}
\usepackage{psfrag}
\usepackage{bm}
\usepackage{epstopdf}
\usepackage{latexsym}
\usepackage{multirow}
\usepackage{amsmath,amssymb,amsfonts}
\usepackage{enumerate}
\usepackage{hyperref}
\usepackage{multirow}
\usepackage{dcolumn}
\usepackage{array}
\usepackage{caption,color}
\def\beq {\begin{equation}}
\def\eeq {\end{equation}}
\def\bea {\begin{eqnarray}}
\def\eea {\end{eqnarray}}
\def\br{\begin{eqnarray}}
\def\er{\end{eqnarray}}
\def\nn {\nonumber}
\def\bc {\begin{center}}
\def\ec {\end{center}}
\def\bi {\begin{itemize}}
\def\ei {\end{itemize}}
\newcommand{\eel}[1] {\label{#1}\end{equation}}

\newcolumntype{L}[1]{>{\raggedright\arraybackslash}p{#1}}
\newcolumntype{C}[1]{>{\centering\arraybackslash}p{#1}}
\newcolumntype{R}[1]{>{\raggedleft\arraybackslash}p{#1}}

\allowdisplaybreaks

\def\d  {\delta}

\def\e  {\epsilon}

\def\lam  {\lambda}
\def\m  {\mu}

\def\o  {\omega}
\def\O  {\Omega}
\def\p  {\pi}

\def \rh  {\rho}

\def\t  {\tau}

\def\la {\langle}

\def \tr {\mbox{Tr}}
\def\f {\frac}

\def\ra {\rangle}

\def\l{\left}
\def\r{\right}
\def\dis{\displaystyle}

\def\til{\tilde}

\begin{document}

\title{Role of  spatial higher order derivatives in momentum space entanglement}
\author{S.\ Santhosh Kumar } \email[email: ]{santhu@iisertvm.ac.in} 
\author{S.\ Shankaranarayanan} \email[email: ]{shanki@iisertvm.ac.in}
\affiliation{\noindent School of Physics, Indian Institute of Science Education and Research Thiruvananthapuram (IISER-TVM), Trivandrum-695016, Kerala, India}
\begin{abstract}
  \noindent We study the momentum space entanglement between different energy modes of 
  interacting scalar fields propagating in general $(D+1)$-dimensional flat space-time.  As opposed 
  to some of the recent  works \cite{Vijay2012-PRD}, we use Lorentz invariant normalized ground state 
 to obtain the momentum space entanglement entropy.   We show that the Lorenz invariant 
 definition removes the spurious power-law behaviour obtained in the earlier works \cite{Vijay2012-PRD}. 
 More specifically, we show that the cubic interacting scalar field in $(1+1)$
  dimensions leads to logarithmic divergence of the entanglement entropy and consistent with the results 
  from real space entanglement calculations.  We study the effects of the introduction of the Lorentz violating higher 
  derivative terms in the presence of non-linear self interacting scalar field potential and show that 
 the divergence structure of the entanglement entropy  is improved in the presence of spatial higher derivative terms.
\end{abstract}
\maketitle
\section{Introduction}
Quantum entanglement depends on two properties ---
the superposition principle and the tensor product structure of the quantum states 
\cite{horodecki2009}. Since the same quantum state has different
tensor product structure in different Hilbert spaces, entanglement entropy
is a partition dependent quantity \cite{zanardi2001,Thirring2011}.  As of now, a large body
of literature has investigated the robustness of the entanglement-area relation for the free quantum 
fields in the real space \cite{bombelli86,srednicki93,cardy2004,shanki2006,eisert2010}. 
It is natural to ask whether the entanglement entropy-area relation gets
modified due to the presence of self interactions.  

In the real space, however, the evaluation of entanglement entropy of self interacting
field runs into difficulty: First, the modes functions can not be evaluated
exactly and, second the evaluation of entanglement entropy is semi-analytical and the validity of the 
numerical results, in the perturbative regime, is opaque. Recently, Balasubramanian et al \cite{Vijay2012-PRD} 
developed a technique to evaluate the entanglement entropy of the self-interacting scalar fields in the 
momentum space.  The procedure of evaluating momentum space entanglement entropy is similar 
to the one used in the evaluation of the real-space entanglement entropy, i. e. the modes of different 
momenta are entangled in the ground state across a particular cut-off which act as
an energy partition in the momentum space. More specifically, the low
energy IR and high energy UV modes are entangled across the cut-off,
say $\m$ \cite{Vijay2012-PRD}.

Like the real space entanglement evaluation, there are unsettled issues in evaluating 
entanglement in the momentum space. First, the entanglement entropy is a cut-off
dependent quantity and still we do not have the correct tool to
renormalize the entropy. Second, the approach has a close resemblance
to the Wilsonian effective low energy action theory as discussed in
Refs.  \cite{WILSON197475,zinn2002quantum,goldenfeld1992lectures}. In
the Wilsonian renormalization, the UV degrees of freedom are
integrated over and IR degrees of freedom are described by an
effective density matrix. This is the case for every interacting field
theory. Third, it is important to note that the momentum space entanglement
is not a universal quantity and depends on what we are integrating
out --- UV or IR modes.  The real space entanglement entropy for a pure 
bipartite system is symmetric w. r. t. the subsystems \cite{Nielsen}. However, it is 
not clear whether the momentum space entanglement entropy satisfies the symmetric
property. The breakdown of symmetric property was reported in
Ref. \cite{Lieb_1963-JMP} for the case of Boson-Fermion duality at
high energy modes.  Hence, the momentum space entanglement entropy is not
useful to characterize theories in an invariant way though, real space
entropy does \cite{Roberts2013-JSTAT}.  It depends on the partitioning
of the UV and IR degrees of freedom. However, in 2-dimensional space-time 
field theories \cite{krishnand2014}, it was shown that the entanglement entropy 
has UV-IR  duality. Our analysis based on the Lorenz invarant definition of the 
ground state, show that the momentum space entanglement entropy is symmetric 
for 2-dimensional field theories. 

As mentioned earlier, the evaluation of the momentum space entanglement was 
first reported in Ref. \cite{Vijay2012-PRD}. While the ground states they used are not
Lorentz invariant, the normalization of these states are not consistent 
with the ones used in the field theory literature~\cite{peskin}. We show that their 
choice of normalization lead to spurious scaling behaviour for the entanglement entropy. 
More specifically, we show that the scaling behavior of the entropy $S  \propto \mu^{D(r-1)-2-r}$ 
used in Ref. \cite{Vijay2012-PRD} go as $S\propto \mu^{D(r-1)-2}$ where $r$ is the 
index of power indicating the strength of the scalar field interaction and $D$ is the number of 
space dimensions.
We explicitly show that the extra power dependence in the entropy reported in Ref. \cite{Vijay2012-PRD} 
is due to their choice of the normalization constant  and show that using the standard 
normalization as in Ref. \cite{peskin}, the results are consistent with real space 
entanglement entropy. We study the effects of the introduction of the Lorentz invariance violating higher 
  derivative terms in the presence of non-linear self interacting scalar field potential and show that 
  the divergence structure of the entanglement entropy  improves when higher derivative terms are taken
into account.

The rest of the paper is organized as follows: Sec. (\ref{sect1})
discusses the approach used to evaluate entanglement entropy in
momentum space and gives explicit formula for calculating
the entanglement entropy of a scalar field in any dimension. In
Sec. (\ref{sect2}), we discuss the model action in any $(D+1)$-
dimensional space-time using the perturbative expansion. In
Sec. (\ref{sect3}), we discuss the results for two specific
cases and generalize to any space dimensions. We show that the 
entropy relation derived here is  different from the ones obtained in 
Ref. \cite{Vijay2012-PRD}  by an extra power factor which has quite compelling 
implications in the renormalization. It is shown that the divergence in the entropy 
is tunable by changing  the dimension of the space-time and the power of the self
interaction. Sec. (\ref{sect4}), concludes with the discussion about
our results and its possible connection to the renormalization of
entanglement entropy. In this work, we set $c = \hbar = 1$. 

\section{Approach to evaluate  momentum space entanglement entropy}
\label{sect1}
Let us start with the first protocol in the pure state entanglement ---
partition the total system into two parts. Let $H_A$ and $H_B$ 
be the Hamiltonian corresponding to two parts $A$ and $B$ with
associated Hilbert spaces $\mathcal H_A$ and $\mathcal H_B$,
respectively. The total Hilbert space of the system is given by, 
\beq
\label{mse1}
\mathcal H=\mathcal H_A\otimes \mathcal H_B 
\eeq 
Let $|n\ra$, $|N\ra$ correspond to the complete energy eigen basis of the subsystem 
$A$ and $B$, respectively.  Furthermore, these represent the occupation numbers of the particle state in the
Fock space. Before switching on the interaction, the Hamiltonian of
the total system is given by, 
\beq
\label{mse2} 
H_0= H_A\otimes \mathcal I_B+\mathcal I_A \otimes H_B \eeq where
$\mathcal I$ is the unit operator and the energy eigenvalues are
represented by $E_n$ and $\til E_N$. The ground state is the tensor
product of the individual ground states of the Hamiltonian $H_A$ and
$H_B$. Mathematically, 
\beq
\label{mse3}  
|0,0\ra \equiv|0_A\ra \otimes |0_B\ra 
\eeq 
After turning on the interaction term, the total Hamiltonian can be written as, 
\beq
\label{mse4}
H=H_0+\lam \,H_{AB} \, .
\eeq 
For the present discussion, we assume that strength of the interaction term 
$H_{AB}$ is weak in order to calculate the interacting ground state ($|\O\ra$) perturbatively. 
Up to first order in $\lam$, we have, 
\br
\label{mse5}
|\Omega\rangle&=&\f{1}{\sqrt{\mathcal N}}\l[ |0,0\rangle +\lambda\left(\dis\sum_{n\neq 0}\frac{\langle n,0|H_{AB}|0,0\rangle}{E_0-E_n}\;|n,0\ra\r.\r.\nn\\
&&\l.\l.\dis+\sum_{N\neq 0}\frac{\langle 0,N|H_{AB}|0,0\rangle}{\tilde E_0-\tilde E_N}\;|0,N\ra\r.\r.\nn\\
&& \l.\l.\dis+\sum_{n,N\neq 0}\frac{\langle n,N|H_{AB}|0,0\rangle}{E_0+\tilde E_0-E_n-\tilde E_N}\;|n,N\ra\right) +\mathcal O(\lambda^2)\r]\nn\\
&=& \f{1}{\sqrt{\mathcal N}}\l[|0,0\rangle +\lambda\left(\displaystyle\sum_{n\neq 0} \mathcal A_n |n,0\rangle+\displaystyle\sum_{N\neq 0}  \mathcal B_N |0,N\rangle\r.\r.\nn\\
&&\l.\l.\qquad\qquad\quad+\displaystyle\sum_{n\neq 0}\sum_ {N\neq 0 } \mathcal C_{n \,N} |n,N\rangle\right) +\mathcal O(\lambda^2)\r]
\er
where 
\br 
\label{mse6}
\mathcal A_n&=&\dis\frac{\langle n,0|H_{AB}|0,0\rangle}{E_0-E_n}, \\
\mathcal B_N&=&\dis\frac{\langle 0,N|H_{AB}|0,0\rangle}{\tilde E_0-\tilde E_N},\\
\mathcal C_{n\,N}&=& \dis\frac{\langle
  n,N|H_{AB}|0,0\rangle}{E_0+\tilde E_0-E_n-\tilde E_N} \er are the
first order coefficients in the perturbative expansion and in general
we can treat them as matrices and $\mathcal N$ is the normalization
constant.

The next protocol is the calculation of total density matrix, the matrix 
entries  are written in the basis of $|0,0\ra , |n,0\ra ,|0,N\ra $ and $ |n,N\ra $ are given by\footnote{For simplicity, we drop the summation,
in the coefficients $\mathcal{A,B}$, $\mathcal C$, and different product combinations.} 
\br
\label{mse7}
\rh&=&|\O\ra\la \O|\\
&=&\dis\f{1}{\mathcal N}\bordermatrix{%
  & \la 0,0| & \la n,0|&\la 0,N|&\la n,N| \cr |0,0\ra & 1&\lam
  \,\mathcal A^\dag &\lam\,\mathcal B^\dag & \lam\,\mathcal C^\dag\cr
  |n,0\ra & \lam\, \mathcal A &\lam^2\, \mathcal {A\,A^\dagger} &
  \lam^2\, \mathcal {A\,B^\dagger} &\lam^2\, \mathcal
  {A\,C^\dagger}\cr |0,N\ra & \lam\,\mathcal B & \lam^2\, \mathcal
  {B\,A^\dagger} & \lam^2\, \mathcal {B\,B^\dagger} &\lam^2\, \mathcal
  {B\,C^\dagger}\cr |n,N\ra & \lam\,\mathcal C &\lam^2\, \mathcal
  {C\,A^\dagger} & \lam^2\, \mathcal {C\,B^\dagger} &\lam^2\, \mathcal
  {C\,C^\dagger} \cr } 
  \er 
  where we fix the normalization constant as,
$\mathcal N=\l(1+|\mathcal A|^2+|\mathcal B|^2+|\mathcal
C|^2\r)^{-1/2}$ and here onwards drop the subscripts attached to
the coefficients $\mathcal {A,B}$ and $\mathcal{C}$.

The reduced density matrix of the $A$ subsystem which is in the basis of $|0\ra $ and $ |n\ra$  is obtained by tracing
over the degrees of freedom of subsystem $B$.  i. e., 
\br
\label{mse8}
\rh_A&=& \dis\tr_B\,\rh\nn\\
&=& \f{1}{\mathcal N}\bordermatrix{ %
  & \la 0| & \la n|\cr |0\ra & 1+\lam^2
  \mathcal{A\,A^\dagger}&\lam\,\mathcal B^\dagger+\lam^2\,
  \mathcal{A\, C^\dagger} \cr |n\ra &\lam\,\mathcal B+\lam^2\,
  \mathcal{C\, A^\dagger}&
  \lam^2\,\l(\mathcal{B\,B^\dagger+C\,C^\dagger}\r)\cr } 
  \er
  The above matrix can be diagonalized and using the unit trace property of the
reduced density matrix, gives the diagonal form of the matrix as, 
\beq
\label{mse9}
\rh_A=\dis\begin{pmatrix}
	1-\lam^2\,|\mathcal C|^2 &0\\
	0& \lam^2\,\mathcal{C C^\dagger}
\end{pmatrix}  +\mathcal O(\lam^3)
\eeq 
The entanglement entropy of the $A$ subsystem is,
\br
\label{mse10}
S_A&=&-\dis\tr_A\l(\rh_A\,\log\rh_A\r)\\
&=& -\tr\l[\l(1-\lam^2\,|\mathcal C|^2\r)\,\log \l(1-\lam^2\,|\mathcal C|^2\r)\r.\nn\\
&&\l.\dis-\tr\l( \lam^2 \mathcal{C\,C^\dagger}\,\log \l(\lam^2 \mathcal{C\,C^\dagger}\r)\r)\r]\\
&\simeq& -\lam^2\log\lam^2 \,\tr \mathcal{\l[C\,C^\dag\r]}\nn\\
&&\dis+\lam^2\, \tr\l[
\mathcal{C\,C^\dagger}\l(1-\log\l[\mathcal{C\,C^\dag}\r]\r)\r]+\mathcal
O(\lam^3) 
\er 
where in the last step we have assumed that $\lam \ll 1$. 
The final expression for the entanglement entropy is, 
\br
\label{mse11}
S_A&=&  -\lam^2\log\lam^2 \, \dis\sum_{n\neq 0}\sum_{ N\neq 0}\frac{|\langle n,N|H_{AB}|0,0\rangle|^2}{\left(E_0+\tilde E_0-E_n-\tilde E_N\right)^2}\nn\\
&& \dis+\lam^2\,\sum_{n\neq 0}\sum_{ N\neq 0}\frac{\langle n,N|H_{AB}|0,0\rangle\langle 0,0|H_{AB}|n,N\rangle}{\left(E_0+\tilde E_0-E_n-\tilde E_N\right)^2}\nn\\
&&\dis \times\l(1-\log\l[\frac{\langle n,N|H_{AB}|0,0\rangle\langle 0,0|H_{AB}|n,N\rangle}{\left(E_0+\tilde E_0-E_n-\tilde E_N\right)^2}\r]\r)\nn\\
&&+\mathcal O(\lam^3) \er 
The leading contribution to entropy is $\lam^2\,(\log \lam^2)$ and vanishes as $\lambda \to 0$.
This is consistent with the fact that the momentum space entanglement entropy of a free field is zero.
In the next section, we apply the above procedure to calculate the momentum
space entanglement entropy for massless interacting scalar field in
different dimensions in the leading order of $\lam^2\,(\log \lam^2)$.

\section{The model: Massless self interacting scalar field}
\label{sect2}

Let us consider the action for a massless scalar field $(\chi)$ propagating in $(D+1)$-dimensional 
flat space-time with linear, higher spatial derivative terms:
\beq
\label{mse12}
\mathcal S=\dis\int dt'\,d^D{\bf y} \left[\frac{1}{2} (\partial_\mu\chi)^2
-\f{\epsilon'}{2}(\nabla^2_{\bf y}\chi)^2-\f{\tau'}{2}(\nabla^3_{\bf y}\chi)^2-\frac{g}{r!} \chi^r\right]
\eeq
where $\e'$ and $\t'$ are the constants with dimensions
[Length]$^{-2}$ and [Length]$^{-4}$ respectively, $ \nabla^2_{\bf y}$
and $ \nabla^3_{\bf y}$ are the higher order spatial derivatives, $g$
is a dimensionfull tunable constant and $r$ refers to the index of
interaction.  The importance of the higher derivative spatial terms
were studied in Refs.  \cite{shanki2012,shanki2016} and can be 
used to understand some quantum phase transitions. 
Unlike the Wilsonian type renormalization \cite{WILSON197475,zinn2002quantum,goldenfeld1992lectures},
the higher derivative terms introduce Next-to-Next-to-Next interaction in the lattice. It is 
interesting to note that these higher derivative terms appear in the effective Hamiltonian 
description of certain high temperature superconductors~\cite{Arturo}.

Rescaling these fields, coupling constants using the following scaling:
\br
\label{mse13}
t'\longrightarrow t=t'/L,  y\longrightarrow x=y/L,&\\
\chi\longrightarrow \phi=\dis\f{\chi}{L^{(1-D)/2}},  \e'\longrightarrow \e=\e' L^{2},&\\
\t'\longrightarrow \t=\t' L^{4}, &\\
 g\longrightarrow \lam= g\,L^{\dis(1+s/2)+D(1-s/2)} \, &
\er
the Hamiltonian $H_{AB}$ corresponding to the action in
eq. (\ref{mse12}) is given by
\br
\label{mse14}
H_{AB}&=&\dis\int d^D{\bf x} \left[\frac{1}{2} \pi^2+\frac{1}{2}
  (\nabla\phi)^2
  +\f{\epsilon}{2}(\nabla^2_{\bf x}\phi)^2\r.\nn\\
  &&\l.\dis+\f{\tau}{2}(\nabla^3_{\bf x}\phi)^2+\frac{\lam}{r!}
  \phi^r\right] \er 
where, $\pi$ is the canonical conjugate momentum corresponding to
scalar field $\phi$ and satisfies the equal time commutation relation
$\l[\phi(\bf x), \pi(x')\r] =i \d^D(\bf x-x')$.

Expanding the field in terms of the bosonic creation and annihilation
operators
\beq
\label{mse15}
\phi({\bf x})=\dis\f{1}{(2\p)^D}\sum_ {\bf p}\f{1}{\sqrt{2 \o_{\bf p}}}\l(a_{\bf p}\, e^{-i\bf p.x}+a^\dag_{\bf p}\, e^{i\bf p.x}\r) \, 
\eeq 
the Lorentz invariant one particle excited states are \cite{peskin,Yu-2004-PRD}:
\beq
\label{mse16} 
|{\bf p}\ra= \sqrt{2{\bf \o_p}}\, a^\dagger_{\bf p}|0\ra
\eeq 
and satisfy Lorentz invariant orthogonality relation \cite{peskin},
\beq
\label{mse17} 
\la {\bf p }|{\bf q}\ra=2\o_{\bf p}\,(2\p)^D \d^D_{\bf p,q}
\eeq  
where $\o_{\bf p}=\dis \sqrt{{\bf p}^2+\e\, {\bf p}^4+\t\, {\bf  p}^6}$. 

We would like to compare and contrast the above relations (\ref{mse16}, \ref{mse17}) to the 
ones used by Balasubramanian et al in Ref. \cite{Vijay2012-PRD}:
\bea
\label{mse16o} 
|{\bf p}\ra=  a^\dagger_{\bf p}|0\ra\\
\la {\bf p }|{\bf q}\ra=(2\p)^D \d^D_{\bf p,q} 
\eea
It is important to note that authors have not used the normalization factor, $\sqrt{2{\bf \o_p}}$ and  this, 
as we will show below, leads to interesting results for the momentum space entanglement entropy.
To understand the effect of the normalization constant, let us calculate  the expectation value of 
$H_{AB}$, i. e.,
\br
\label{mse18}
& \!\!\!\!\!\!\!\!\!\!\!\!\!\!\!\!\!\!\!\!\!\!\!\! \!\!\!\!\!\!\!\!\!\!\!\!\!\!\!\!\!\!\!\!\!\!\!\!\!\!\!\!\!\!\!\!\! \dis\la {\bf p_1,p_2\ldots, p_r}|H_{AB}|0,\ldots,0\ra  = \nn\\
& \dis\la {\bf p_1,p_2\ldots, p_r}|\int d^D{\bf x}\,\f{ \phi({\bf x})^r}{r!}|0,0,\ldots,0\ra \nn\\
=& \!\!\!\!\!\!\!\!\!\!\!\!\!\!\!\!\!\!\!\!\!\!\!\!\!\!\!\!\!\! \dis\la {\bf p_1,p_2\ldots, p_r}|\int \f{d^D{\bf x}}{r!}\,\l(\dis\f{1}{(2\p)^D}\r.\nn\\
&\l.\dis\times\sum_ {\bf k}\f{1}{\sqrt{2 \o_{\bf k}}}\l(a_{\bf k}\, e^{-i\bf k.x}+a^\dag_{\bf k}\, e^{i\bf k.x}\r)\r)^r|0,\ldots,0\ra\nn\\
= &\dis \sum_{\bf k_1,\ldots k_r}\f{1}{(2\p)^{D\,r}r!}\int d^D{\bf x}\, e^{i\l(\bf k_1+k_2+\ldots+k_r\r).{\bf x}}\nn\\
& \dis \times\f{\la \bf p_1,p_2\ldots, p_r| k_1,k_2\ldots, k_r\ra}{(2\o_{\bf k_1}2\o_{\bf k_2}\ldots 2\o_{\bf k_r})}\nn\\
=& \!\!\!\!\!\!\!\!\!\!\!\!\!\!\!\!\!\!\!\!\!\!\!\! \!\!\!\!\!\!\!\!\!\!\!\!\!\!\!\!\!\!\!\!\!\!\!\!\!\!\!\!\!\!\!\!\! \dis \f{(2\p)^D}{r!} \d^D_{{\bf p_1+ p_2+\ldots+ p_r}}
\er     
where in the last step we have used equations (\ref{mse16}) and (\ref{mse17}) simultaneously. The above 
expression is different from Eq. (29) of Ref. \cite{Vijay2012-PRD}, importantly, it does not contain the inverse 
frequency terms. 

The leading term in the momentum space entanglement entropy is then given by:
{\small
\br
\label{mse19}
S_A&=&  -\lam^2\log\lam^2 \, \dis\sum_{\{\bf p_1,\ldots p_r\}_\m}\frac{(2\p)^{2D}}{(r!)^2}\,\f{\d^D_{{\bf p_1+ p_2+\ldots+ p_r}}}{\l(\o_{\bf p_1}+\o_{\bf p_2}+\ldots +\o_{\bf p_r}\r)^2}\nn\\
&&\dis+\mathcal O(\lam^2)
\er
}
\noindent where the summation is over the momentum such that at least one
momentum is below the cut off $\m$ and at least one momentum above the
cut off \cite{Vijay2012-PRD}. More specifically, we study the entanglement between the high and low energy modes across the cut-off $\mu$. 
Here, we are interested to understand the momentum space entropy of the low energy mode by tracing over the high energy modes 
above the cut-off like in the Wilsonian effective action. Taking the continuum limit, i.e, 
\br 
\sum_{\bf p}\longrightarrow\f{1}{(2\p)^D}\int d^D{\bf
  p}, &\qquad\d_{\bf p}\longrightarrow (2\p)^D \d^D({\bf p}) 
\er
Eq. (\ref{mse19}) becomes: 
\br
\label{mse20}
S_A&=&  -\lam^2\log\lam^2 \, \prod_{j=1}^r\dis\int_{\{\bf p_1,\ldots p_r\}_\m} d^D {\bf p}_j \frac{(2\p)^{2D}}{(r!)^2}\,\nn\\
&&\dis\times\f{\d^D({{\bf p_1+ p_2+\ldots+ p_r}})}{\l(\o_{\bf p_1}+\o_{\bf p_2}+\ldots +\o_{\bf p_r}\r)^2}+\mathcal O(\lam^2)
\er
Let us compare and contrast Eq. (\ref{mse20}) with Eq. (31) of Ref. \cite{Vijay2012-PRD} that is reproduced here 
for easy comparison:
\br
\label{mse201}
S_A/L^D&=&  -\lam^2\log\lam^2 \prod_{j=1}^r\, \dis\int_{\{\bf p_1,\ldots p_r\}_\m} \frac{d^D {\bf p}_j}{(2\p)^{D(r-1)} \,2^r(r!)^2}\,\nn\\
&&\dis\times\f{\d^D({{\bf p_1+ p_2+\ldots+ p_r}})}{\o_{\bf p_1}\ldots .\o_{\bf p_r}\l(\o_{\bf p_1}+\ldots +\o_{\bf p_r}\r)^2}+\mathcal O(\lam^2)
\er
First, the entanglement entropy is a dimensionless quantity. However, Eq.~(31) of Ref. \cite{Vijay2012-PRD} is a dimensionfull 
quantity.   Second, the  consideration of the Lorentz invariant ground state removes the  factors,  $\o_{\bf p_1}\ldots .\o_{\bf p_r}  $ from the 
denominator of Eq. (\ref{mse201}). In the next subsection, we obtain the momentum space entanglement of the interacting fields 
with higher spatial derivative terms.

\section{Role of spatial higher derivatives in momentum space entanglement}
\label{sect3}

Having shown the importance of the normalization constant in the evaluation of the momentum 
space entanglement entropy, in this section, we discuss the role the spatial higher derivatives 
play in the divergence of the entanglement entropy.  We evaluate the entanglement entropy 
for two specific examples --- $\phi^3$ theory in $(1+1)$ and $\phi^4$ theory in $(2+1)$ dimensions 
--- and generalize to generic interaction in arbitrary dimensional space-time.

\subsection{$\phi^3$ theory in (1+1)-dimensions}

Setting $r=3$ and $D=1$ in Eq. (\ref{mse20}),  entanglement entropy is given by:
\br
\label{mse21}
S_A&=&  -\lam^2\log\lam^2 \, \dis\int_{\{ p_1,p_2, p_3\}_\m} d { p_1} d { p_2} d { p_3} \frac{(2\p)^{2}}{(3!)^2}\nn\\
&&\times \dis\f{\d({{ p_1+ p_2+ p_3}})}{\l(\o_{ p_1}+\o_{ p_2} +\o_{ p_3}\r)^2}+\mathcal O(\lam^2) \, ,
\er
where $\omega_{\bf p} = \sqrt{b \, {\bf p}^2+\e\, {\bf p}^4+\t\, {\bf  p}^6}$ and $b$ takes values $0$ or $1$.
Evaluating the above integrals, such that at least one momentum is below the cut off $\m$ and at least 
one momentum is above the cut off \cite{Vijay2012-PRD} and using the Lorentz invariant orthogonality relation
(\ref{mse17}), the leading order term in the entanglement entropy (in the large $\mu$ limit) is
\beq
\label{mse22}
S_A \propto -\lambda^2 \log \lambda^2 
\begin{cases}
 \log \m , ~\mbox{for}\, \epsilon = \tau = 0;b = 1 \\
	\dis\f{1}{\mu^2}, ~  \mbox{for} \, b = \tau = 0; \e=1 \\ 
	\dis\f{1}{\mu^4}, ~\mbox{for}\, b = \epsilon = 0; \t=1 
\end{cases} +\mathcal O(\lam^2)
\eeq
This is one of the key results of this work regarding which we would to stress the following points:
First, in the absence of the higher derivative spatial terms $\epsilon = \tau = 0$, the entanglement 
entropy depends on the cutoff logarithmically and not a power-law. This is consistent with the 
analysis in Ref. \cite{krishnand2014}. It was shown that the divergence of the entanglement entropy of $(1 + 1)-$
dimensional field theory can be mapped to IR problem and should be valid for small values of the coupling 
constant $\lambda$. Thus, the results obtained here for the momentum space entanglement entropy are consistent. 
Second, the UV-IR mapping of 2-dimensional field theories \cite{krishnand2014} also shows that the momentum space entanglement entropy 
with the correct normalization constant is symmetric  w.r.t. partitioning. Third, introduction of the higher derivative terms 
improve the divergence structure of the entanglement entropy i. e. 
$\nabla^2$ term leads to entanglement entropy decaying as $\mu^2$. The divergence structure of quantum field theory is 
expected to vastly improve when higher derivative terms are taken into account.  In particular, introducing $\Box^2$ term 
to the scalar field theory leads to logarithmic divergence instead of power-law divergence of the two-point function~\cite{1983-Barth.Christensen-PRD}.
Thus, our analysis is consistent with these results.

\subsection{$\phi^4$ theory in (2+1)-dimensions}

Let us now consider $\phi^4$ interacting, massless scalar field propagating in $(2+1)$-dimension. Substituting $ D = 2$ and 
$r = 4$, in Eq. (\ref{mse20}), the momentum space entanglement entropy in the large $\mu$ limit is given by
\beq        
\label{mse24}
S_A\propto -\lambda^2 \log \lambda^2 
\begin{cases}
	\dis \m^4, ~~\mbox{for} \, \e = \tau = 0; b=1\\
	\dis \m^2, ~~ \mbox{for}\, b = \tau = 0; \e=1 \\
	\dis\log\m, ~~ \mbox{for} \, b = \epsilon = 0; \t=1
\end{cases}+\mathcal O(\lam^2)
\eeq 
As in the previous case, our results show that the introduction of the higher derivative 
makes the entanglement entropy less divergent and is consistent with the analysis of 
Ref. \cite{1983-Barth.Christensen-PRD}. 

In general for any self interacting scalar model of the type in Eq.  (\ref{mse14}) in D -dimensional space, 
the momentum space entanglement entropy, in the large $\mu$ limit, is given by 
\beq      
\label{mse23}
S_A\propto -\lambda^2 \log \lambda^2 
\begin{cases}
	\dis \f{1}{\m^{2-D(r-1)}}, ~~ \mbox{for} \, \e=\tau = 0; b=1\\
	\dis \f{1}{\dis \m^{4-D(r-1)}},~~ \mbox{for}\,  b = \tau = 0; \e=1 \\ 
	\dis\f{1}{\m^{6-D(r-1)}}, ~~ \mbox{for}\,  b = \e = 0; \t=1 
\end{cases}+\mathcal O(\lam^2)
\eeq
The above results show that the interaction terms do not improve the 
divergence problem of the entanglement entropy, however, the higher 
derivative terms improve the divergence structure of the entanglement entropy. 

\section{Conclusions}
\label{sect4}

In this work, we have evaluated the momentum space entanglement entropy of 
the interacting scalar fields in the presence of spatial higher derivative terms. 
We have explicitly shown that the correct choice of the normalization of the Lorenz 
invariant states does not improve the divergence problem of the entanglement 
entropy, however, the presence of higher derivative terms improves the divergence 
of the entanglement entropy.

Our analysis should be contrasted to the analysis reported in Ref. \cite{Vijay2012-PRD}
where the authors claimed that the interaction terms help to improve the divergence 
problem of entanglement entropy.  As we have shown here, this is due to the wrong 
choice of the normalization constant. Taking two specific examples, scalar fields in 
$(1+1)-$ dimensions and $(2 + 1)-$ dimensions, we have shown that the divergence 
structure of the entanglement entropy is not improved due to the presence of the 
interaction terms. 

Real space entanglement entropy is symmetric, however, the momentum space 
entanglement defined in Ref. \cite{Vijay2012-PRD} is not and hence 
can not be considered as an universal quantity~\cite{Roberts2013-JSTAT}. Our analysis 
in Sec. (IV-A), in the light of Ref. \cite{krishnand2014},  shows that at least for the 
$(1 + 1)-$dimensional field theory, the momentum space entanglement entropy can {\sl 
indeed} be considered as an universal quantity. More specifically, since the UV and IR 
are related by a simple rescaling of the variables in $( 1 + 1)-$dimensions~\cite{krishnand2014}, 
the momentum space entanglement entropy evaluated by integrating over the UV 
modes or IR modes is identical.  Our aim is to extend the analysis for higher 
dimensions. This is currently under investigation.

\section{Acknowledgments}
 SSK is financially supported by Senior Research Fellowship of the Council of Scientific \& Industrial Research (CSIR), Government of India.  SSK is also thankful to the Condensed Matter and
 Statistical Physics section at ICTP, Italy for the hospitality
 during the final stages of the manuscript. This work
 is supported by Max Planck-India Partner Group on Gravity and
 Cosmology.  
 
 \providecommand{\href}[2]{#2}\begingroup\raggedright\endgroup


\begin{thebibliography}{10}
 	
 	\bibitem{Vijay2012-PRD}
 	V.~Balasubramanian, M.~B. McDermott, and M.~Van~Raamsdonk, ``Momentum-space
 	entanglement and renormalization in quantum field theory,"
 	\href{http://dx.doi.org/10.1103/PhysRevD.86.045014}{{\em Phys. Rev. D}
 		{\bfseries 86}  045014 (2012)}.
 	
 	\bibitem{horodecki2009}
 	R.~Horodecki, P.~Horodecki, M.~Horodecki, and K.~Horodecki, ``Quantum
 	entanglement,'' \href{http://dx.doi.org/10.1103/RevModPhys.81.865}{{\em Rev.
 			Mod. Phys.} {\bfseries 81}  865--942 (2009)}.
 	
 	\bibitem{zanardi2001}
 	P.~Zanardi, ``Virtual quantum subsystems,''
 	\href{http://dx.doi.org/10.1103/PhysRevLett.87.077901}{{\em Phys. Rev. Lett.}
 		{\bfseries 87}  077901 (2001)}.
 	
 	\bibitem{Thirring2011}
 	W.~Thirring, R.~A. Bertlmann, P.~K{\"o}hler, and H.~Narnhofer, ``Entanglement
 	or separability: the choice of how to factorize the algebra of a density
 	matrix,'' \href{http://dx.doi.org/10.1140/epjd/e2011-20452-1}{{\em The
 			European Physical Journal D} {\bfseries 64} no.~2,  181--196 (2011)}.
 	
 	\bibitem{bombelli86}
 	L.~Bombelli, R.~K. Koul, J.~Lee, and R.~D. Sorkin, ``Quantum source of entropy
 	for black holes,'' \href{http://dx.doi.org/10.1103/PhysRevD.34.373}{{\em
 			Phys. Rev. D} {\bfseries 34}  373--383 (1986)}.
 	
 	\bibitem{srednicki93}
 	M.~Srednicki, ``Entropy and area,''
 	\href{http://dx.doi.org/10.1103/PhysRevLett.71.666}{{\em Phys. Rev. Lett.}
 		{\bfseries 71}  666--669 (1993)}.
 	
 	\bibitem{cardy2004}
 	P.~Calabrese and J.~L. Cardy, ``{Entanglement entropy and quantum field
 		theory},'' \href{http://dx.doi.org/10.1088/1742-5468/2004/06/P06002}{{\em
 			Journal of Statistical Mechanics: Theory and Experiment}\\ {\bfseries 0406}
 		 P06002 (2004)}.
 	
 	\bibitem{shanki2006}
 	S.~Das and S.~Shankaranarayanan, ``How robust is the entanglement entropy-area
 	relation?,'' \href{http://dx.doi.org/10.1103/PhysRevD.73.121701}{{\em Phys.
 			Rev. D} {\bfseries 73}  121701 (2006)}.
 	
 	\bibitem{eisert2010}
 	J.~Eisert, M.~Cramer, and M.~B. Plenio, ``\textit{Colloquium} : Area laws for
 	the entanglement entropy,''
 	\href{http://dx.doi.org/10.1103/RevModPhys.82.277}{{\em Rev. Mod. Phys.}
 		{\bfseries 82}  277--306 (2010)}.
 	
 	\bibitem{WILSON197475}
 	K.~G. Wilson and J.~Kogut, ``The renormalization group and the $\epsilon $
 	expansion,''
 	\href{http://dx.doi.org/http://dx.doi.org/10.1016/0370-1573(74)90023-4}{{\em
 			Physics Reports} {\bfseries 12} no.~2,  75 -- 199 (1974)}.
 	
 	\bibitem{zinn2002quantum}
 	J.~Zinn-Justin, {\em ``Quantum Field Theory and Critical Phenomena,"}
 	\newblock International series of monographs on physics, Clarendon Press, (2002).
 	
 	\bibitem{goldenfeld1992lectures}
 	N.~Goldenfeld, {\em ``Lectures on Phase Transitions and the Renormalization
 		Group,"}
 	\newblock Frontiers in physics, Addison-Wesley, Advanced Book Program, (1992).
 	
 	\bibitem{Nielsen}
 	M.~A. Nielsen and I.~L. Chuang, {\em ``Quantum Computation and Quantum
 		Information,"}
 	\newblock Cambridge University Press, 10th anniversary edition~ed., (2011).
 	
 	\bibitem{Lieb_1963-JMP}
 	D.~C. Mattis and E.~H. Lieb, ``Exact solution of a many-fermion system and its
 	associated boson field,''
 	\href{http://dx.doi.org/http://dx.doi.org/10.1063/1.1704281}{{\em Journal of
 			Mathematical Physics} {\bfseries 6} (1965) }.
 	
 	\bibitem{Roberts2013-JSTAT}
 	M.~Headrick, A.~Lawrence, and M.~Roberts, ``Bose-fermi duality and entanglement
 	entropies,'' \href{http://dx.doi.org/10.1088/1742-5468/2013/02/P02022}{{\em
 			Journal of Statistical Mechanics: Theory and Experiment} \\{\bfseries 2013}
 		no.~02,  P02022 (2013)}.
 	
 	\bibitem{krishnand2014}
 	K.~Mallayya, R.~Tibrewala, S.~Shankaranarayanan, and T.~Padmanabhan, ``Zero
 	modes and divergence of entanglement entropy,''
 	\href{http://dx.doi.org/10.1103/PhysRevD.90.044058}{{\em Phys. Rev. D}
 		{\bfseries 90}  044058 (2014)}.
 	
 	\bibitem{peskin}
 	M.~Peskin and D.~Schroeder, {\em ``An Introduction to Quantum Field Theory,"}
 	\newblock Advanced Book Program, Addison-Wesley Publishing Company, (1995).
 	
 	\bibitem{shanki2012}
 	S.~Ghosh and S.~Shankaranarayanan, ``Entanglement signatures of phase
 	transition in higher-derivative quantum field theories,''
 	\href{http://dx.doi.org/10.1103/PhysRevD.86.125011}{{\em Phys. Rev. D}
 		{\bfseries 86}  125011 (2012)}.
 	
 	\bibitem{shanki2016}
 	S.~S. Kumar and S.~Shankaranarayanan, ``{Evidence for phase transition in
 		vacuum entanglement of higher derivative scalar quantum field theories},''
 	\href{http://arxiv.org/abs/1606.05472}{{\em arXiv:1606.05472
 			[cond-mat.stat-mech]} (2016)}.
 	
 	\bibitem{Arturo}
 	A.~Tagliacozzol, 
 	\newblock Private communication.
 	
 	\bibitem{Yu-2004-PRD}
 	Y.~Shi, ``Entanglement in relativistic quantum field theory,''
 	\href{http://dx.doi.org/10.1103/PhysRevD.70.105001}{{\em Phys. Rev. D}
 		{\bfseries 70}  105001 (2004)}.
 	
 	\bibitem{1983-Barth.Christensen-PRD}
 	N.~H. Barth and S.~M. Christensen, ``Quantizing fourth order gravity theories.
 	1. the functional integral,''
 	\href{http://dx.doi.org/10.1103/PhysRevD.28.1876}{{\em Phys. Rev. D.}
 		{\bfseries 28}  1876 (1983)}.
 	
 \end{thebibliography}
\end{document}